\documentclass[aps,pra,twocolumn,noeprint,superscriptaddress]{revtex4-1}
\usepackage{amsmath,amssymb,amsfonts,bbm,graphicx,times,color}
\usepackage{subfigure}%
\usepackage{mwe}
\usepackage[breaklinks=true,colorlinks=true,linkcolor=blue,urlcolor=blue,citecolor=blue]{hyperref}
\newcommand{\ket}[1]{\vert #1 \rangle}

\newcommand{\ketbra}[2]{\vert #1 \rangle \langle #2 \vert}
\def\imm{\mathrm{i}}

\begin{document}
\title{Quantum state engineering by non-deterministic 
noiseless linear amplification}
\author{Hamza Adnane}\email{hamza.adnane92@gmail.com}
\affiliation{Laboratoire de Physique Th\'eorique, Universit\'e de B\'eja\"ia,
Campus Targa Ouzemour, 06000 B\'eja\"ia, Algeria}
\author{Matteo Bina}\email{matteo.bina@gmail.com}
\affiliation{Quantum Technology Lab, Dipartimento di Fisica, Universit\`a degli Studi di
Milano, I-20133 Milano, Italy}
\author{Francesco Albarelli}\email{francesco.albarelli@warwick.ac.uk}
\affiliation{Department of Physics, University of Warwick, 
Coventry CV4 7AL, United Kingdom}
\author{Abdelhakim Gharbi}\email{hakimgharbi74@gmail.com}
\affiliation{Laboratoire de Physique Th\'eorique, Universit\'e de B\'eja\"ia,
Campus Targa Ouzemour, 06000 B\'eja\"ia, Algeria}
\author{Matteo G. A. Paris}\email{matteo.paris@fisica.unimi.it}
\affiliation{Quantum Technology Lab, Dipartimento di Fisica, Universit\`a degli 
Studi di Milano, I-20133 Milano, Italy}
\date{\today}
\begin{abstract}
We address quantum state engineering of single- and two-mode 
states by means of non-deterministic noiseless linear amplifiers 
(NLAs) acting on Gaussian states. In particular, we show that 
NLAs provide an effective scheme to generate highly non-Gaussian 
and non-classical states. Additionally, we show that the amplification 
of a two-mode squeezed vacuum state (twin-beam) may highly increase 
entanglement.
\end{abstract}
\maketitle
\section{Introduction}
Non-classical and non-Gaussian states of continuous variable systems 
have been a relevant resource in the development of quantum information 
science and technology, as well as in several fundamental 
tests of quantum mechanics itself \cite{rev1,rev2,rev3}. Quantum 
state engineering of those states, however, is often hampered by 
two major challenges. On the one hand, generation of nonclassical light 
usually involves nonlinear optical media, and the small 
value of nonlinear susceptibilities leads to low efficiency.
On the other hand, light amplification is usually involved as 
well, and the linear and unitary character of quantum dynamics makes 
this task rather difficult, since it imposes that noise should 
be unavoidably added when a signal is amplified, in order to maintain 
the uncertainty relation \cite{caves}.
\par
Recently, in order to circumvent these difficulties, 
novel amplifying devices have been suggested, which act 
non-deterministically, i.e. the output state is obtained conditionally by post-selecting on a particular measurement outcome \cite{ral09,xia10,wal13,fer12,fer11,mar10,Combes2016}.
In particular, optimal schemes describing non-deterministic
linear amplifiers (NLAs) achieving successful amplification 
with the largest probability allowed by
quantum mechanics have been put forward~\cite{Pandey2013,mcm14}.
This kind of devices are appealing for quantum state engineering,
especially in the continuous-variable regime, where several
schemes based on conditional states of continuous variables measurements 
have been already explored and proved effective~\cite{ral03,kni01,ral05,col03,koz96,lau03,Genoni2010a,Arzani2017}.
\par
In this paper, we address quantum state engineering of single- 
and two-mode states by means of non-deterministic noiseless 
linear amplifiers acting on Gaussian states. In particular, we 
prove that NLAs provide an effective scheme to generate highly 
non-Gaussian  and non-classical states. We provide a general
framework to address state engineering by NLA, and present
explicit results for the non-Gaussianity and non-classicality
obtained with single-mode coherent and squeezed vacuum 
states, as well as with two-mode squeezed vacuum (twin-beam
state). In order to assess quantitatively the performances of
NLAs on those signals, we quantify non-classicality using negativity
of the Wigner function (from now on W-nonclassicality) \cite{ken04} and non-Gaussianity by the relative entropy to a reference Gaussian state \cite{m1,m2}.
We remark that these and other closely related quantifiers have recently been studied also in the context of quantum resource theories~\cite{Zhuang2018a,Albarelli2018,Takagi2018,Park2018}.
\par
Concerning the action of the NLA on two-mode states, we consider 
two working regimes. In the first one, we analyse the performances 
of a strong (destructive) NLA measurement on twin-beam and 
analyse the properties of the resulting single-mode conditional 
states \cite{col03}.  Our results show that depending on the 
squeezing parameter and the gain of the device, the resulting 
conditional state has a Wigner 
function that includes negative parts. This suggests that such a scheme 
represents a robust source for non-classicality. The second regime 
that we are going to explore is that of non-destructive NLA 
measurement on twin-beam. Here, we address the degree of entanglement 
of the amplified state as a function of the NLA gain parameter, and 
prove that it may be significantly enhanced.
\par
The paper is organized as follows. In Sec~\ref{s:pre}, we establish notation
and review the non-Gaussianity and non-classicality measures used across 
this work, that is, the entropic non-Gaussianity and the W-nonclassicality.
We also recall the main ingredients needed to describe non-deterministic 
linear amplifiers. In sec~\ref{s:one}, we  discuss quantum state engineering 
by NLA on coherent and squeezed vacuum states, focussing our attention on non-Gaussianity and 
non-classicality of the conditionally amplified states. 
Sec~\ref{s:two} is devoted to conditional states generation by exploiting
the action of NLAs on twin-beam, in both the destructive and 
non-destructive regimes. Section \ref{s:out} closes the paper with
concluding remarks.
%%%%%%%%%%%%%%%%%%%%%%%%%%%%%%%%%
\section{Preliminaries}\label{s:pre}
In this Section, we briefly introduce the tools we will use throughout 
the paper to quantify non-classicality (nC) and non-Gaussianity (nG)
of a single-mode bosonic quantum state. For the characterization of 
nG we employ a measure based on the quantum relative entropy (QRE) 
between the state under examination and a reference Gaussian state 
\cite{m1,m2}.
Concerning the nC, we use an indicator based on the volume of the negative part of the Wigner function~\cite{ken04}.
The final Subsection is devoted to briefly review the description of a noiseless linear amplifier in terms of measurement operators of to introduce the corresponding conditionally amplified states~\cite{mcm14}.
\subsection{Non-Gaussianity based on Quantum Relative Entropy} 
\label{s:nG}
In continuous-variable systems, a single-mode radiation field is 
described by creation and annihilation operators, $\hat{a}^{\dagger}$ \ 
and \ $\hat{a}$ respectively, satisfying the bosonic commutation 
relation $[\hat{a},\hat{a}^{\dagger} ]=\hat{\mathbb{I}}$.
Coherent states are the eigenstates of the annihilation operator
$\ket{\alpha}$, such that $\hat{a}\ket{\alpha}=\alpha\ket{\alpha}$, 
whereas number states $\ket{n}$ are the eigenstates of the number operator $\hat{a}^\dag \hat{a}\ket{n}=n\ket{n}$. 
The operators $\hat{x}=(\hat{a}+\hat{a}^\dag)/\sqrt{2}$ and 
$\hat{p}=\imm (\hat{a}^\dag -\hat{a})/\sqrt{2}$, describe the
observable  quadratures of the field.
A quantum state, i.e. a density operator $\hat{\varrho}$, 
may be represented in the 
phase space by means of the characteristic function 
$\chi[ \hat{\varrho}]  (\alpha)=\text{Tr}[ \hat{\varrho} 
\hat{D}(\alpha)]  ,$ where $\hat{D}(\alpha)=\exp\{  
\alpha \hat{a}^{\dagger}-\alpha^* \hat{a} \}  $ \ is 
the displacement operator. 
The Wigner function, defined as the Fourier transform of the 
characteristic function $W[\hat{\varrho}](\lambda)\propto\int 
\text{d}^{2} \alpha\exp\{\lambda^*\alpha-\lambda\alpha^*\}
\chi[\hat{\varrho}](\alpha)$, is the most iconic quasi-probability 
distribution for the quantum state~\cite{schleich2001quantum}.
In particular, it is the only one in the family of the $p$-ordered quasiprobability distributions, widely used in quantum optics~\cite{Cahill1969a,Barnett1997}, that gives the probability densities for quadrature measurements as its marginal distributions. 
\par
Gaussian states~\cite{fer05,ger14,oli12,serafini2017quantum} are quantum states 
having a Gaussian Wigner function 
\begin{equation} \label{WGaussian}
W[\hat{\varrho}_G] (\mathbf{X})=\frac{\exp \big [-\frac12 (\mathbf{X}-\langle \hat{\mathbf{R}}\rangle)^T\boldsymbol{\sigma}^{-1}(\mathbf{X}-\langle \hat{\mathbf{R}}\rangle)\big ]}{2\pi \sqrt{\text{det}\,\boldsymbol{\sigma}}} \, 
\end{equation}
where we considered the Cartesian representation of real
variables $\boldsymbol{X}=(x,p)^T$.
A Gaussian state is fully identified by its first-moment vector and it's 
covariance matrix (CM), given by
\begin{subequations}\begin{align}
\langle \hat{\mathbf{R}}\rangle&=\big (\langle\hat{x}\rangle,\langle\hat{p}\rangle \big) \label{Rave}\\
[\boldsymbol{\sigma}]_{kl}&=\frac{1}{2}\langle \{  \hat{R}_{k},\hat{R}_{l}\}  \rangle -\langle \hat{R}_{k}\rangle \langle \hat{R}_{l}\rangle \, , \label{CM}
\end{align}\end{subequations}
where $\hat{\mathbf{R}}\equiv (\hat{x},\hat{p})$, the anti-commutator is denoted as $\{ \cdot\, , \cdot \}  $  and the expectation values $\langle \cdot \rangle $  are calculated over $\hat \varrho$ by the Born rule.
\par
A useful measure of nG for a quantum state may be obtained by 
introducing a reference Gaussian state $\hat \varrho_{G}$, having 
the same CM and first-moment vector of state under consieration 
$\hat{\varrho}$, namely $\langle \hat{\mathbf{R}}_G\rangle =
\langle \hat{\mathbf{R}}\rangle$ and $\boldsymbol{\sigma}_G = 
\boldsymbol{\sigma}$.
Given the von Neumann entropy 
$S(\hat{\varrho})=-\text{Tr}[\hat{\varrho}\ln\hat{\varrho}]$, 
the nG measure is then defined as 
the quantum relative entropy (QRE) of these two states, i.e. 
$\delta_\mathrm{nG}[\hat{\varrho}] \equiv S[\hat{\varrho} \,|\!| \hat{\varrho}_{G}]=\text{Tr}[\hat{\varrho}(\ln\hat{\varrho}-\ln\hat{\varrho}_{G})] $, 
which, eventually, reduces to
\begin{equation}\label{nG}
  \delta_\mathrm{nG}[\hat{\varrho}]=S(\hat{\varrho}_G)-S(\hat{\varrho}) \, ,
\end{equation}
by exploiting the assumptions made on the reference 
Gaussian state $\hat{\varrho}_G$ \cite{m1}.
Furthermore, this quantifier corresponds to the relative entropy of nG, i.e. the minimum relative entropy between the state under consideration and the whole set of Gaussian states~\cite{Marian2013a}.

The quantum relative entropy is not a proper metric, since it is not symmetric under exchange of its arguments. In spite of this issue, it has been profitably used as a measure of statistical distinguishability between quantum states~\cite{ved02}. 
The von Neumann entropy of a single mode Gaussian state 
is fully determined by its CM as $S( \hat{\varrho}_{G}) =
h\left(  \sqrt{\det\mathbf{\boldsymbol{\sigma}}}\right)$
where the function $h(x)$ is given by
\begin{equation}
h(x)=\left (x+\frac{1}{2}\right )\ln\left (x+\frac{1}{2}\right )
-\left( x-\frac{1}{2}\right )\ln\left ( x-\frac{1}{2}\right ).
\label{7}
\end{equation}
In the following, we will be mostly interested in pure 
states, for which $S(\hat{\varrho})=0$; the 
nG measure thus assumes a simple form
\begin{equation}\label{nGpure}
 \delta_\mathrm{nG}[\hat{\varrho}]=h\left(  \sqrt{\det\mathbf{\boldsymbol{\sigma}}}\right)  \, .
\end{equation}
%%%
\subsection{Non-classicality based on Wigner Negativity } \label{s:nC}
Phase-space analysis is at the heart of several approaches to detect 
and quantify the non-classical character of quantum states. Different 
notions of nC may be introduced, stemming from the quasi-probability 
distributions associated to the state under scrutiny, or by minimizing 
the distance to a set of classical states, similarly to the nG measurement 
we have just introduced.
\par
From the physical point of view, the most relevant notion of 
non-classicality is associated to the Glauber-Sudarshan $P$-function.
A state is $P$ non-classical when its $P$-function is not a 
well-behaved density of  probability, i.e., includes singularities 
and/or negative parts \cite{gla63,man86,pnc12}.
This notion of nonclassicality stems from the fact that 
coherent states are the only pure quantum states that 
show classical features from the point of view of quantum optics.
A good measure of $P$-nC is the so-called 
{\em non-classical depth}~\cite{le1,le2}, which consists in 
evaluating the minimal amount of Gaussian noise required to 
turn the $P$-function into a well-behaved probability distribution. 
A caveat of this measure is that all pure nG states saturate the upper
bound of the non-classical depth, i.e. they are all maximally 
non-classical under this measure~\cite{lut95}. As a consequence 
this measure is not helpful to analyze in details the nC properties 
of the pure Gaussian states we are going to consider hereafter.
This notion of nonclassicality has been recently studied in the 
context of resource theories~\cite{Tan2017}, where an operational 
interpretation in terms of metrological usefulness has been 
introduced~\cite{Kwon2018,Yadin2018}.
\par
An alternative approach, based on Wigner function, has been suggested
few years ago~\cite{ken04}. The W-function is known to be a well-defined 
quasi-probability density function,  i.e., it may display negative 
parts, but it never has singularities. The $W$-nonclassicality ($W$-nC) 
corresponds to the negative volume of the Wigner function, and it provides 
an intuitive and practical way to detect the amount of nC of a quantum 
state, as it allows to distinguish the degree of nC for different 
pure states. Moreover, this weaker form of non-classicality (any $W$ 
non-classical state is also $P$ nonclassical, while the opposite is 
not necessarily true) is the crucial resource for several quantum 
information tasks (see below).
\par
$W$-nC for a generic quantum state $\hat{\varrho}$ is defined 
as follows:
\begin{equation}\begin{split}
\delta_\mathrm{nC}[\hat{\varrho}]&=\int \text{d} x \,\text{d}p\big[  \big\vert W[\hat{\varrho}](x,p)\big\vert -W[\hat{\varrho}](x,p)\big]  \\
&=\int \text{d} x \,\text{d}p \big\vert W[\hat{\varrho}](x,p)\big\vert   -1\, , \label{deltaNC}
\end{split}\end{equation}
where the integration is performed over the whole phase space and 
the second equality is obtained taking into account the normalization 
of the Wigner function. Notice that the Wigner function may also be computed 
directly, without passing through the characteristic function, upon employing 
the following expression, valid for any density operator
\begin{equation}\label{Wgen}
W[\hat{\varrho}](\boldsymbol{X})=\frac{2}{\pi}\text{Tr}\left [ \hat{\varrho} \hat{D}(2 \boldsymbol{X}) \hat{\Pi} \right ] \, ,
\end{equation}
with the parity operator defined as $\hat{\Pi}=(-1)^{\hat{a}^\dag \hat{a}}$.
\par
From an operational point of view, the notion of W-nC has been connected 
to the impossibility of efficiently simulating a quantum system via 
phase-space methods~\cite{Mari2012,Veitch2013}, or more quantitatively 
to the hardness of estimating the output probabilities of an experiment~\cite{Pashayan2015}. In turn, this fact seems to play an important role 
for schemes aimed at quantum supremacy with homodyne detection~\cite{Douce2016,Chakhmakhchyan2017}. For these reasons, W-nC has been 
studied in the context of an operational resource theory where the 
free operations are Gaussian ones; a particularly useful monotone 
is the so-called Wigner logarithmic negativity~\cite{Albarelli2018}, 
defined as $\log \left ( \delta_\mathrm{nC}[\hat{\varrho}] + 1\right)$.
\par
We also stress that the connection between nG states and nC states with 
a negative Wigner function is strong. According to the Hudson 
theorem~\cite{Hudson1974}, Gaussian states are the only \emph{pure} 
states with a positive $W$, while for mixed states the situation 
is more involved~\cite{Brocker1995a,Mandilara2009,Hughes2014}.
Interestingly, for families of pure non-Gaussian states where only 
a single parameter is varied, quantifiers of nG and $W$-nC were 
always found to be in a monotonic relationship.
In particular, this observation has been made for ground states 
of anharmonic oscillators~\cite{Albarelli2016a}, where the 
behaviour of the two quantities is also qualitatively very similar.
However, note that in some cases the behaviour can be wildly 
different, while retaining monotonicity, e.g. when varying 
the amplitude of cat states the measure $\delta_\mathrm{nC}$ 
saturates to a finite value, while $\delta_\mathrm{nG}$ 
diverges~\cite{Albarelli2018}.
\par
Notice also that the quantifier of W-nC in Eq.~(\ref{deltaNC}) 
has the valuable property of being accessible through experimental 
measurements, since the Wigner function can be reconstructed 
by means of tomographic techniques, involving photon counting 
or homodyne detection of the marginal distributions of 
$W[\hat{\varrho}](x,p)$ \cite{ban99,vog00,lvo02,zam09}.
Another experimentally friendly quantifier of $W$-nC based 
on relative entropy was also proposed~\cite{Mari2011}.
\par
\begin{figure}[h!]
\center
\includegraphics[width=0.48\textwidth]{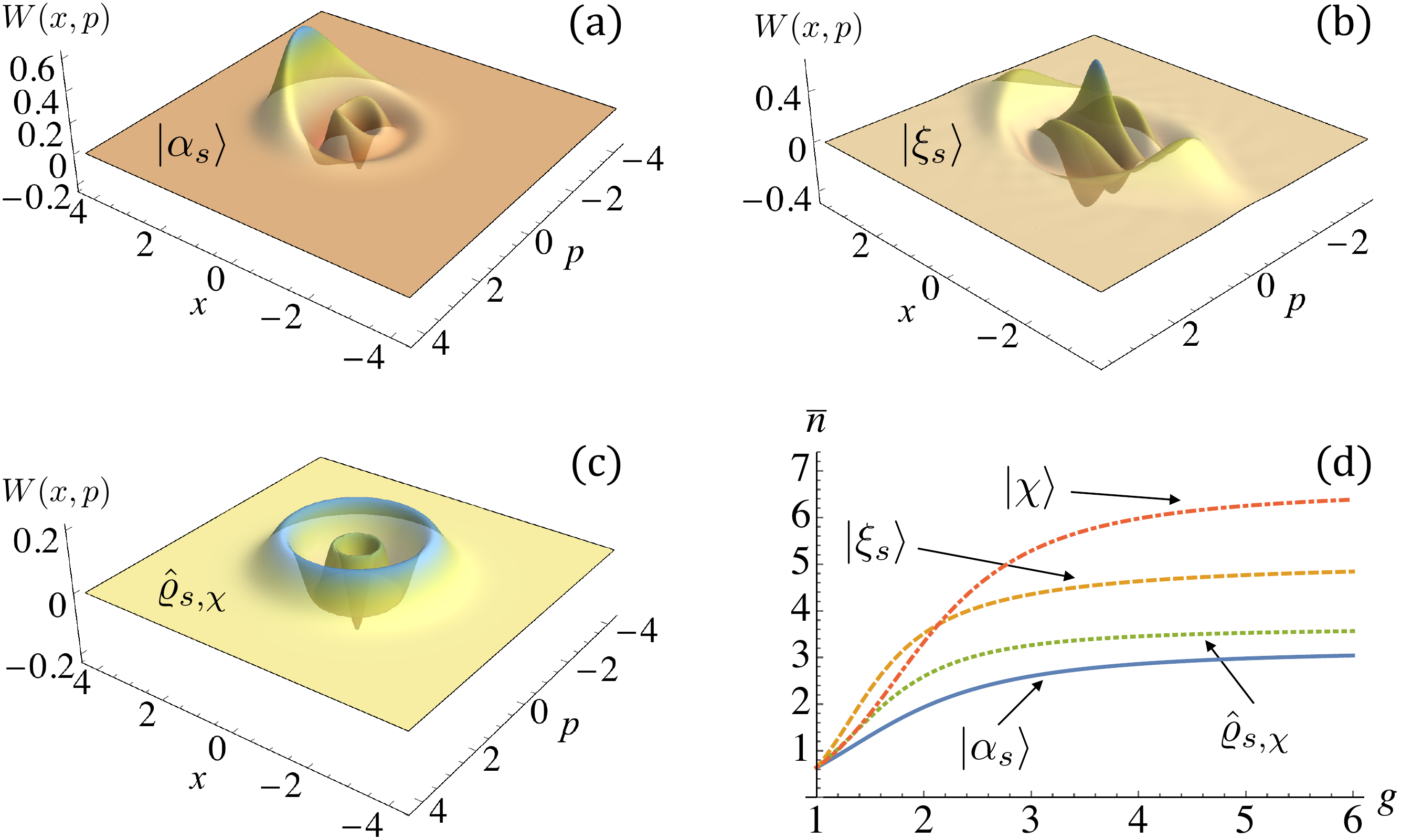}
\caption{(Color online) Wigner functions $W(x,p)$ of (a) the amplified coherent state $\ket{\alpha_s}$ of Eq.~(\ref{Coherent_s}), (b) the amplified squeezed vacuum state $\ket{\xi_s}$ of Eq.~(\ref{SqVacAmp}) and (c) the reduced state of an amplified twin-beam $\hat{\varrho}_{s,\chi}$ of Eq.~(\ref{TWBredAmp}). These amplified states clearly show a highly non-classical and non-Gaussian behavior. The plots have been obtained by setting: (a) $\alpha=0.8$, (b) $\xi=0.73$, (c) $\chi=0.63$ and $g=4$, $p=3$. (d) The effect of amplification in the average photon number $\bar{n}$ for these states, together with the amplified TWB of Eq.~(\ref{TWBamp}), is plotted against the gain parameter $g$, assuming the same average photon number $\bar{n}=0.64$ at $g=1$. } \label{f:Wfunctions}
\end{figure}

In the following Sections, we are going to consider different kinds of 
signals undergoing noiseless amplification. In order to qualitatively 
anticipate the nature of our results, let us provide a
{\em phase-space snapshot} of the corresponding conditionally amplified 
states: in Fig.~\ref{f:Wfunctions} we show the Wigner functions of an
amplified coherent state, an amplified squeezed vacuum state, and the 
reduced state of an amplified twin-beam. The $W$-negativity and the nG
character of the amplified states clearly emerge. We also show the effect
of amplification in terms of the output signal energy (i.e. average photon 
number), assuming the same average photon number  $\bar{n}$ at $g=1$.
%%%
\subsection{Noiseless linear amplification}
Ideally, a perfect amplification would be obtained by applying the operator $\hat{T}=g^{\hat{a}^\dag\hat{a}}$, with a positive gain parameter $g>1$.
The idea behind the NLA is to approximate the action of this ideal operator, by implementing a measurement protocol with post-selection~\cite{ral09}.
However, it is not possible to implement the ideal operation with a finite probability of success and therefore there is a trade-off between the probability of success and the fidelity to the desired results.
The usual approach to obtain a feasible implementation with non-zero probability of success is to implement the action of the ideal amplification only on a finite-dimensional truncation of the Fock space.
An optimal (w.r.t. the previously mentioned trade-off) measurement (Kraus) operator for this scheme was introduced in~\cite{Pandey2013} and further studied~\cite{mcm14}.
The action of this protocol is described as follows: we consider two possible outcomes, 
success ($s$) and failure ($f$), which corresponds to a 
two-element POVM $\left\{  \hat{E}_{s}^{p\dagger}\hat{E}_{s}^{p},\text{ }
\hat{E}_{f}^{p\dagger}\hat{E}_{f}^{p}\right\}$ such that $\hat{E}_{s}^{p\dagger}\hat{E}_{s}^{p}+\hat{E}_{f}^{p\dagger}\hat{E}_{f}^{p}=\hat{\mathbb{I}}$. 
The success measurement operator reads
\begin{equation}
\hat{E}_{s}^{p}=g^{-p}\sum\limits_{n=0}^{p}g^{n}\left\vert n\right\rangle \left\langle n\right\vert +\!\!\sum\limits_{n=p+1}^{\infty}
\left\vert n\right\rangle \left\langle n\right\vert \, ,
\end{equation}
where $p$ is an integer truncation parameter of the Fock space, i.e. the
{\em amplification threshold}. The probability of a successful amplification 
when the measurement scheme is applied to a generic pure state 
$\left\vert \psi\right\rangle $ may be written as
\begin{align}
P_{s,\psi} & =\langle \psi\vert \hat{E}_{s}^{p\dagger}\hat{E}_{s}^{p}\vert 
\psi\rangle \notag \\  & 
=g^{-2p}\!\sum\limits_{n=0}^{p} g^{2n}\vert\langle n | \psi\rangle 
\vert ^{2}+\!\!\sum\limits_{n=p+1}^{\infty}\!\vert \langle n | \psi\rangle \vert ^{2}\,. 
\end{align}
The corresponding conditionally amplified state is
\begin{equation}\label{CondState}
\left\vert \psi_{s}\right\rangle =\frac{\hat{E}_{s}^{p}\left\vert
\psi\right\rangle }{\sqrt{P_{s,\psi}}}\,.
\end{equation}
In the limit of low gain, i.e. $g=1+\gamma$ with $0< \gamma \ll 1$ we may
write the success operator in the following simplified form
\begin{align}
\hat{E}_{s}^{p} \stackrel{\gamma \ll 1}{\simeq} {\mathbb I} - 
\gamma \sum_{n=0}^p\, (p-n)\, |n\rangle\langle n|\,.
\end{align}
The corresponding probability of amplification may be written as 
\begin{align}
P_{s,\psi} \stackrel{\gamma \ll 1}{\simeq} 1 - 2 \gamma \sum_{n=0}^p\, 
(p-n)\, \big|\langle\psi |n\rangle\big|^2\,.
\end{align}

In the following, we will consider amplification of 
paradigmatic examples of Gaussian states, with emphasis on the 
conditional generation of non-Gaussian and non-classical amplified 
states by means of the NLA process. 
\section{Engineer non-Gaussian and non-classical states by noiseless amplification 
of single-mode signals}\label{s:one}

\subsection{Noiseless amplification of coherent states}\label{s:coh}
In order to assess the performances of NLAs in the generation on nC and nG, 
let us start by investigating their action on coherent states, which can 
be easily generated experimentally.
Coherent states correspond to displaced vacuum states $\hat{D}(\alpha)\ket{0}$,  and may be expressed in the Fock basis as
\begin{equation}\label{Coherent}
\ket{\alpha} ={\rm e}^{-\frac{| \alpha| ^{2}}{2}}
\sum_{n=0}^{\infty}
\frac{\alpha^{n}}{\sqrt{n!}}\ket{ n } , 
\end{equation}
where $\alpha$ is the coherent state amplitude 
and $\bar{n}=\left\vert \alpha\right\vert ^{2}$ the mean photon number.
According to Eq.~(\ref{CondState}), an amplified coherent state 
reads 
\begin{equation}\label{Coherent_s}
\ket{\alpha_{s}} =\frac{e^{-\frac{| \alpha |^{2}}{2}}}{\sqrt{P_{s,\alpha}}}\left(  g^{-p}
\sum_{n=0}^{p}\frac{(g\alpha)^{n}}{\sqrt{n!}}\ket{ n} +\sum_{n=p+1}^{\infty} \frac{\alpha^{n}}{\sqrt{n!}}\ket{ n} \right)  , 
\end{equation}
where $P_{s,\alpha}$ denotes the probability of a successful amplification
\begin{equation}
P_{s,\alpha}=e^{-| \alpha | ^{2}}\left[  g^{-2p}
\sum_{n=0}^{p}
\frac{(g|\alpha|)^{2n}}{n!}+
\sum_{n=p+1}^{\infty}
\frac{|\alpha|^{2n}}{n!}\right]  .
\end{equation}
\par
\begin{figure}[h!!]
\center
\includegraphics[width=0.48\textwidth]{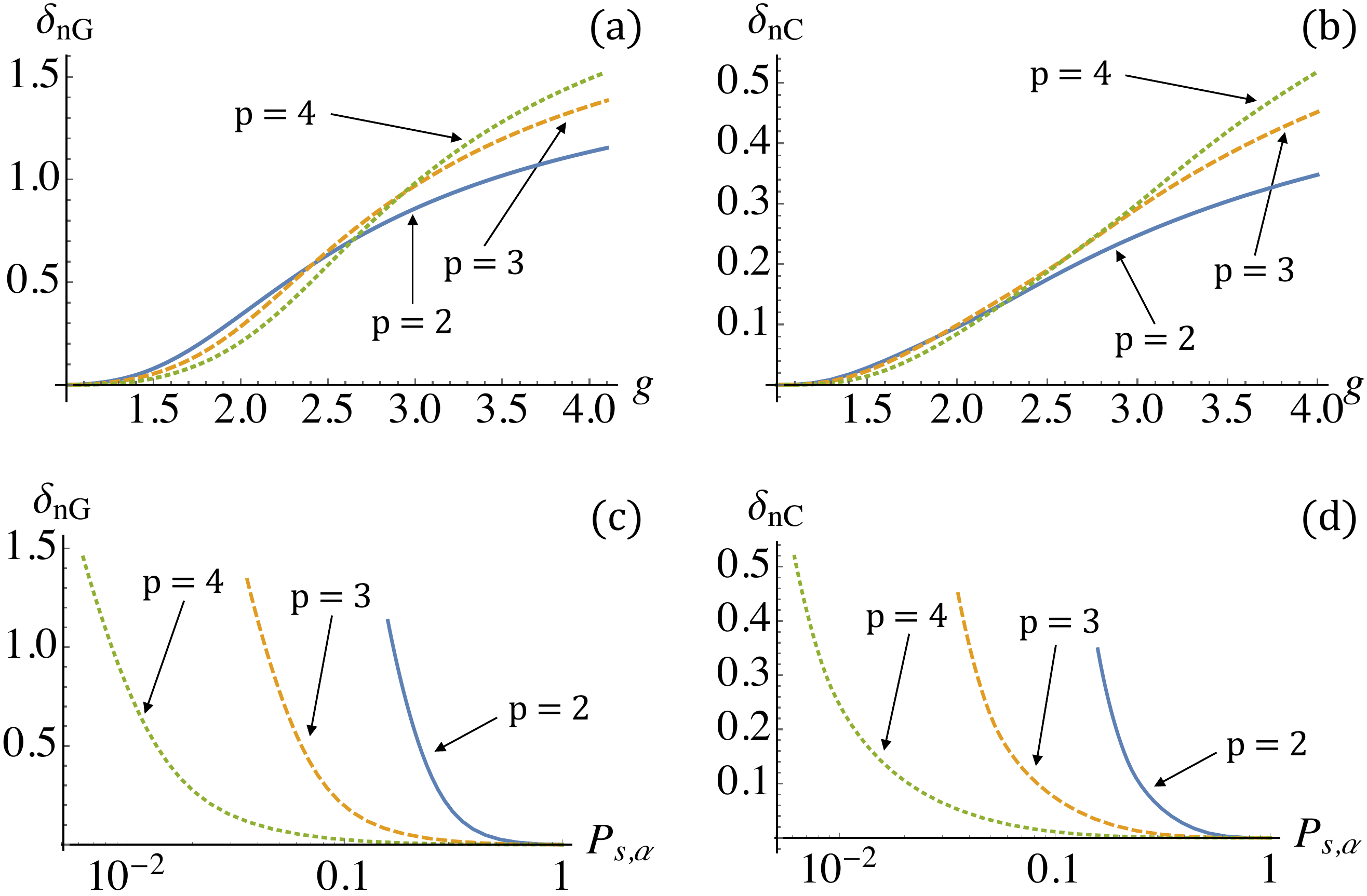}
\caption{(Color online) Non-Gaussianity and non-classicality of the 
amplified coherent state of Eq.~(\ref{Coherent_s}). (a) Plot of the nG measure $\delta_\mathrm{nG}$ as a function of the gain parameter $g$. (b) Plot of the nC measure $\delta_\mathrm{nC}$ as a function of the gain parameter $g$. Plots of $\delta_\mathrm{nG}$ (c) and $\delta_\mathrm{nC}$  (d) as a function of the success probability $P_{s,\alpha}$. The three curves for each plot correspond to different NLA 
thresholds: $p=2$ (solid blue), $p=3$ (dashed orange) and 
$p=4$ (dotted green). The amplitude of the coherent state 
is $\alpha=0.8$.} \label{f:nGnCCoherent}
\end{figure}

The nG of the resulting amplified coherent state is assessed in terms of the measure $\delta_\mathrm{nG}$, as described in Sec.~\ref{s:nG}, and, since 
the conditional state $\ket{\alpha_{s}}$ generated by the NLA
is pure, it can be easily computed by means of Eq.~(\ref{nGpure}). 
As shown in the upper left panel of Fig.~\ref{f:nGnCCoherent}, 
the action of the non-deterministic NLA on a coherent state is
to generate an amplified noiseless pure state (\ref{Coherent_s}) 
with an amount of nG monotonically increasing with the gain 
parameter $g$. In particular, we chose, without loss of generality, 
a real coherent amplitude $\alpha=0.8$ and different values of 
the threshold parameter $p=2,3,4$. As it is apparent from the plot, 
we have a monotone behaviour with $g$ for any $p$. The larger is $p$
the larger is the nG at large values of the gain.
%%%
\par
The Wigner function of the amplified coherent state 
$\hat{\varrho}_{s,\alpha}=\ketbra{\alpha_s}{\alpha_s}$ 
may be conveniently obtained from 
the expression (\ref{Wgen}). As 
it can be appreciated from the plot in Fig.~\ref{f:Wfunctions}(a), 
the Wigner function clearly has negative values.
In the upper right panel of Fig.~\ref{f:nGnCCoherent}, the $W$-nC of an
 amplified coherent state, i.e. $\delta_\mathrm{nC}[\hat{\varrho}_{s,\alpha}]$ 
 in Eq.~\ref{deltaNC}, is shown as 
 a function of the gain $g$ of the NLA. We notice that $W$-nC increases 
 monotonically with $g$, and the larger is the threshold, the larger is the nC, 
 for large $g$. Remarkably, both the nG and the nC share the same qualitative 
 behaviour against the NLA parameters, confirming the connection explained in~\ref{s:nC}.  
 The behaviours of the two quantities 
 suggest to identify the NLA gain as the parameter 
 driving the semiclassical coherent signal to a highly non-Gaussian and 
 non-classical one. The lower panels of Fig.~\ref{f:nGnCCoherent} illustrate 
 the tradeoff between the amount of obtained nG (nC) and the corresponding 
 probability $P_{s,\alpha}$ of successful amplification. Larger values 
 of $P_{s,\alpha}$ are obtained for smaller gain.
\subsection{Noiseless amplification of squeezed vacuum}
\begin{figure}[h!]
	\center
	\includegraphics[width=0.48\textwidth]{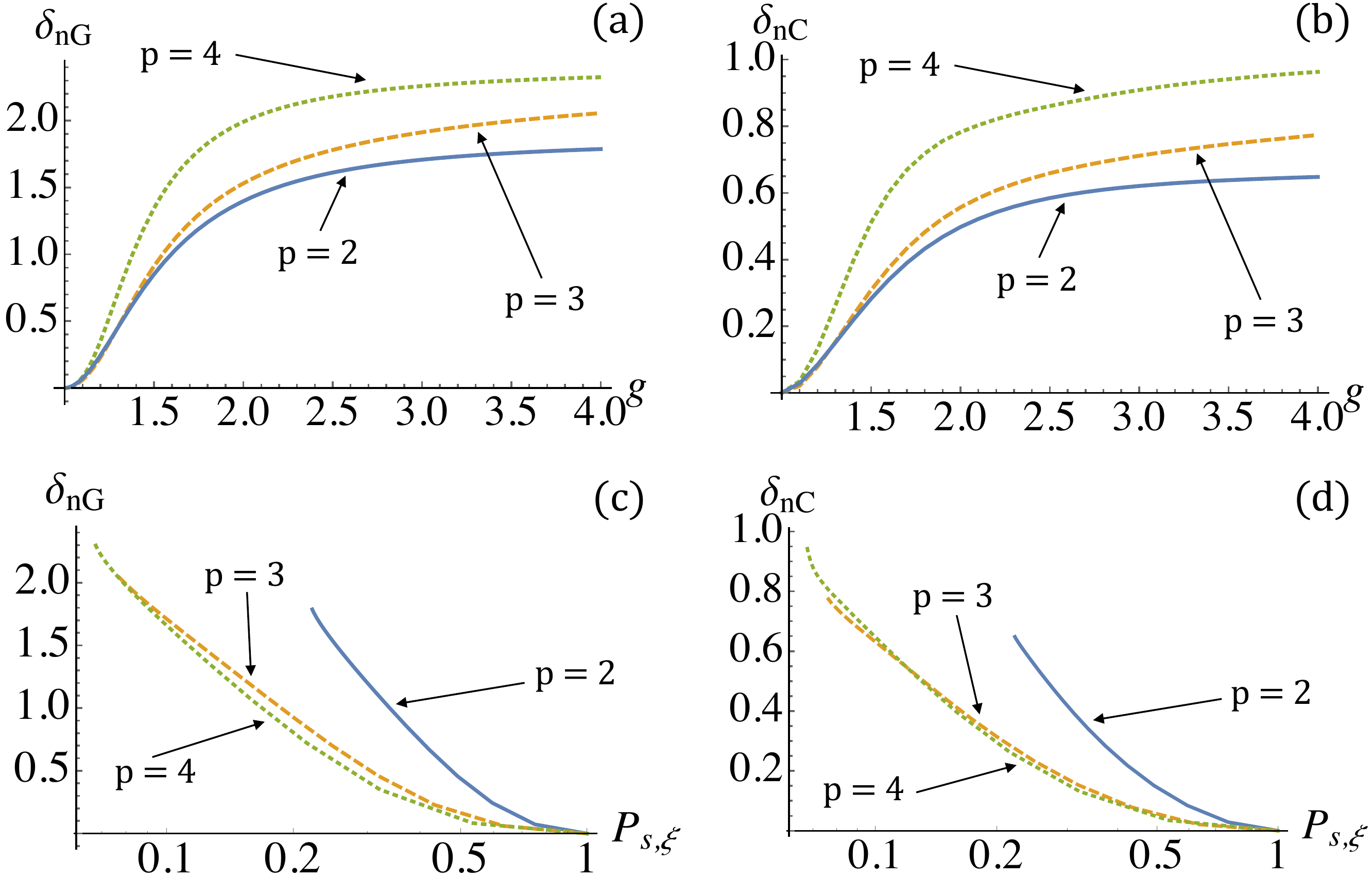}
\caption{(Color online) Non-Gaussianity and non-classicality of the 
amplified squeezed vacuum state of Eq.~(\ref{SqVacAmp}). 
(a) nG measure $ \delta_\mathrm{nG} $ plotted against the gain 
parameter $g$. (b) Plot of the nC measure $\delta_\mathrm{nC}$ as a 
function of the gain $ g $. Plots of $\delta_\mathrm{nG}$ (c) 
and $\delta_\mathrm{nC}$  (d) as a function of the success 
probability $P_{s,\xi}$. The three curves for each plot
correspond to different NLA threshold: $p=2$ (solid blue), 
$p=3$ (dashed orange) and $p=4$ (dotted green). The squeezing 
parameter is set to $r=0.73$.
	}	
	\label{f:nGnCSqueezed}
\end{figure}

Another important Gaussian state, employed as a resource 
in many quantum protocols, is the squeezed vacuum 
state $\ket{\xi}=\hat{\mathcal{S}}(\xi)\ket{0}$, where the 
squeezing operator $\hat{\mathcal{S}}(\xi)=\exp\{\frac{1}{2}
\xi(\hat{a}^{\dagger})^{2}-\frac{1}{2}\xi^{\ast}(\hat{a})^{2}\}$ 
acts on the vacuum state. The phase of the squeezing parameter 
$\xi=re^{i\phi}$ specifies which quadrature of the field is squeezed, 
whereas its modulus quantifies the amount of squeezing. The expression 
of a squeezed vacuum in the Fock basis is given by
\begin{align}
\ket{\xi} & =\frac{1}{\sqrt{\mu }}\sum_{n=0}^{\infty}\left(  
\frac{\nu}{2 \mu}\right)  ^{n}\frac{\sqrt{(2n)!}}{n!}\ket{2n} 
\notag \\ & \equiv\sum_{n=0}^\infty x_n\,\ket{2n}\,, \quad \sum_{n=0}^\infty |x_n|^2 =1 
\end{align}
where $ \mu=\cosh r$ and $\nu=e^{i\phi}\sinh r$.
The action of NLA on the squeezed vacuum with a successful 
amplification, employing Eq.~(\ref{CondState}), reads
\begin{equation}\label{SqVacAmp}
\ket{\xi_{s}} = \frac{1}{\sqrt{P_{s,\xi}}}\left ( g^{-p}\sum_{\substack{n=0 \\  (n \text{ even})}}^p g^n x_{\frac{n}{2}}\ket{n} + \sum_{\substack{n=p+1 \\ (n \text{ even})}}^\infty x_{\frac{n}{2}}\ket{n} \right ) \, ,
\end{equation}
where the success probability is given by
\begin{equation}
P_{s,\xi}=g^{-2p}\sum_{n=0}^p g^{2n}|x_{\frac{n}{2}}|^2+
\sum_{n=p+1}^\infty |x_{\frac{n}{2}}|^2 \, .
\end{equation}
In the upper left panel of Fig.~\ref{f:nGnCSqueezed}, we show the 
nG measure  $\delta_\mathrm{nG}[\hat{\varrho}_{s,\xi}]$ for the amplified state 
$\hat{\varrho}_{s,\xi}=\ketbra{\xi_s}{\xi_s}$ as a function of 
the gain parameter $g$, for different values of the threshold $p=2,3,4$, 
at a fixed value of the squeezing parameter $\xi=0.73$, corresponding 
to the same input energy, at $g=1$, of a coherent state with $\alpha=0.8$. 
We notice that, starting from a squeezed vacuum, the efficiency of the NLA 
in generating non-Gaussianity is much larger than in the 
coherent-state case, in particular for low values of the 
gain parameter.
\par
In the upper right panel of Fig.~\ref{f:nGnCSqueezed}, we show the
non-classicality  $\delta_\mathrm{nC}[\hat{\varrho}_{s,\xi}]$ as a function 
of the gain parameter $g$, for different values of the threshold $p=2,3,4$, 
at a fixed value of the squeezing strength $r=0.73$. Similarly to the 
nG measure, amplification generates a large amount of non-classicality, 
also for low values of the gain. The role of the threshold $p$, 
likewise, is to increment the 
nC measure. 
Again, $\delta_\mathrm{nG}[\hat{\varrho}_{s,\xi}]$ and 
$\delta_\mathrm{nC}[\hat{\varrho}_{s,\xi}]$ are two increasing 
monotonic functions, and the amplified state is highly non-classical and 
non-Gaussian at the same time.
The lower panels of Fig.~\ref{f:nGnCSqueezed} illustrate 
the tradeoff between the amount of nG (nC) obtained and 
the success probability $P_{s,\alpha}$. Remarkably, 
increasing the threshold value from $p=3$ to $p=4$ leads 
to substantially larger values of nG and nC, with basically 
the same success probability. 
\section{Engineer non-Gaussian and non-classical states by noiseless 
amplification of twin-beam}\label{s:two}
\subsection{Destructive noiseless amplification of twin-beam}\label{s:sec4}
The projection postulate offers a viable mechanism to realise synthetic
dynamics and, in turn, to generate quantum states otherwise unreachable 
with Hamiltonian evolution \cite{col03}. In this Section, we analyse the 
effects of both destructive and non-destructive implementation of NLA
on a maximally entangled continuous variables states, namely the 
twin-beam (TWB) state:
\begin{equation}
\left\vert \chi\right\rangle =\sqrt{1-\chi^{2}}\, \sum\limits_{n=0}^{\infty}
\chi^{n}\left\vert n , n\right\rangle , \label{TWB}
\end{equation}
where $0<\chi<1$ and $\ket{n,n}=\ket{n}\otimes\ket{n}$ is the Fock basis for the 
two-mode system. The TWB state is a Gaussian two-mode state obtained by the 
action of the two-mode squeezing operator $\hat{\mathcal{S}}_2(\chi)=
\exp\{\chi \hat{a}\hat{b}-\chi^*\hat{a}^\dag\hat{b}^\dag\}$ on the vacuum 
$\ket{0}\otimes\ket{0}$, where $a$ and $b$ denote the two involved modes.
Without loss of generality, we will consider 
$\chi$ as a real parameter. These states of light may be generated in 
non-degenerate optical parametric amplifiers by spontaneous down-conversion 
or by mixing at a balanced linear mixer two squeezed vacua with opposite 
squeezing phases. Since the efficiencies of these processes are relatively 
weak, it is of interest to investigate protocols to enhance the resulting 
nonclassical properties.
\par
According to the reduction postulate, a measurement performed on one of 
the two constituents of an entangled bipartite system leaves the other 
part in a conditional state, which depends on the outcome of the 
measurement. Considering the twin-beam state (\ref{TWB}) and a successful 
NLA amplification performed on subsystem $a$, the subsystem $b$ is 
reduced into a diagonal state
\begin{align}
\hat{\varrho}_{s,\chi} & =\frac{1}{P_{s,\chi}}\text{Tr}_{a}\left[  \left\vert \chi\right\rangle
\left\langle \chi\right\vert \hat{\Pi}_{s}\otimes \hat{\mathbb{I}}_b\right]\\
&=\frac{(1-\chi^{2})}{P_{s,\chi}}\Bigg[  g^{-2p}\,\sum\limits_{n=0}^{p}(g\chi)^{2n}\left\vert n\right\rangle \left\langle n\right\vert \notag \\ & \qquad 
\quad \qquad \qquad 
+\sum_{n=p+1}^{\infty}\chi^{2n}\left\vert n\right\rangle \left\langle n\right\vert
\Bigg]  , \label{TWBredAmp}%
\end{align}
where $\hat{\Pi}_{s}=\hat{E}_{s}^{p\dagger}\hat{E}_{s}^{p}$ denotes 
an element of the POVM describing the NLA, $\hat{\mathbb{I}}_{b}$ the
identity operator acting on $\mathcal{H}_{b}$ and $P_{s,\chi}$ is the 
probability of successful amplification:
\begin{equation} \label{Pschi}
P_{s,\chi}=(1-\chi^{2})\left[  g^{-2p}
{\displaystyle\sum\limits_{n=0}^{p}}
(g\chi)^{2n}+\sum_{n=p+1}^{\infty}\chi^{2n}\right] \, .
\end{equation}
In order to compute the nG measure for the amplified state 
in $\hat{\varrho}_{s,\chi}$ we refer to Eq.~(\ref{nG}). As the 
state (\ref{TWBredAmp}) is diagonal in the Fock basis, its 
reference Gaussian state $\hat{\varrho}_G$ is a thermal 
state with mean photon number $\bar{n}_{s,\chi}=
\text{Tr}[\hat{\varrho}_{s,\chi}\,\hat{a}^\dag\hat{a}]$ 
and diagonal CM $\boldsymbol{\sigma}=\frac12\text{Diag}
(1+2\bar{n}_{s,\chi},1+2\bar{n}_{s,\chi})$. Moreover, the 
von Neumann entropy $S(\hat{\varrho}_{s,\chi})=-\sum_n
\rho_n\ln\rho_n$ can be directly calculated with the 
diagonal matrix elements $\rho_n$ of the state (\ref{TWBredAmp}). 
The resulting nG measure for this amplified mixed state can, 
thus, be written in this simple form:
\begin{equation}\label{nGredTWB}
\delta_\mathrm{nG}[\hat{\varrho}_{s,\chi}]=h\left (\frac12+
\bar{n}_{s,\chi}\right) + \sum_n\rho_n\ln\rho_n\,.
\end{equation}
\begin{figure}[h!]
	\center
	\includegraphics[width=0.48\textwidth]{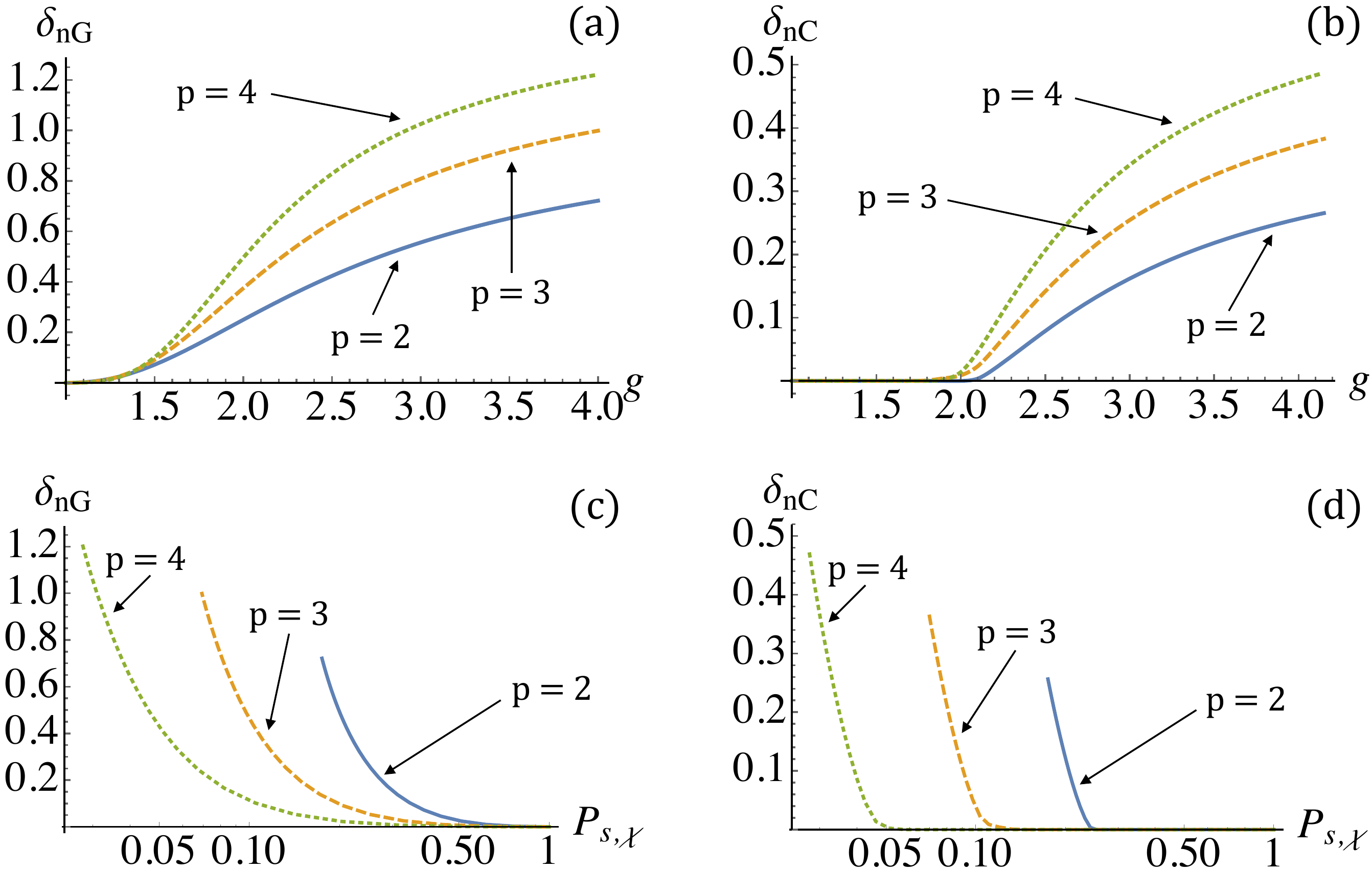}
	\caption{(Color online)  Non-Gaussianity and non-classicality 
	of the amplified state of Eq.~(\ref{TWBredAmp}). (a) nG measure 
	$ \delta_\mathrm{nG} $ plotted against the gain parameter $g$. (b) plot 
	of the nC measure $\delta_\mathrm{nC}$ as a function of the gain $ g $. 
	Plots of $\delta_\mathrm{nG}$ (c) and $\delta_\mathrm{nC}$  (d) as a function 
	of the success probability $P_{s,\chi}$.The three curves for each 
	plot correspond to different thresholds $p=2$ (solid 
	blue), $p=3$ (dashed orange) and $p=4$ (dotted green). The TWB 
	parameter is set to $\chi=0.63$.
	} 
	 \label{f:TWBred}
\end{figure}
\par
In the upper left panel of Fig.~\ref{f:TWBred} we show the nG measure (
\ref{nGredTWB}) as a function of the gain parameter $g$ for the 
amplified state $\hat{\varrho}_{s,\chi}$. The three curves correspond 
to different thresholds $p=2,3,4$ and, as already observed in the 
previous cases, the function $\delta_\mathrm{nG}[\hat{\varrho}_{s,\chi}]$ 
increases monotonically with the gain. The same happens if one 
increases the threshold.
\par
As highlighted in Fig.~\ref{f:Wfunctions}(c), the Wigner function 
of the amplified state $\hat{\varrho}_{s,\chi}$ may assume negative values.
The nC measure $\delta_\mathrm{nC}[\hat{\varrho}_{s,\chi}]$ is shown 
as a function of the gain $g$ in the upper right panel of 
Fig.~\ref{f:TWBred} and it displays a monotonic growth. 
At variance with the previous cases, i.e. amplification of 
coherent and squeezed vacuum states, we observe that the W-nC of 
$\hat{\varrho}_{s,\chi}$ is zero before a threshold value of 
the gain, and then starts to grow monotonically for any value of $p$.
This happens since the Wigner function of the amplified state is 
diagonal in the Fock basis and, thus, phase independent. It 
begins to warp as the gain increases but remains positive. Only 
after a particular value of the gain, depending mainly on the TWB 
parameter $\chi$, the Wigner function takes on negative volumes.
Thus we have found that the two quantifiers we considered are still in a monotonic relationship, even though they have a qualitatively different behavior, as already mentioned in Sec.~\ref{s:nC}.
%%\subsubsection{Comparison of performances}
\par
\begin{figure}[h!]
	\center
	\includegraphics[width=0.245\textwidth]{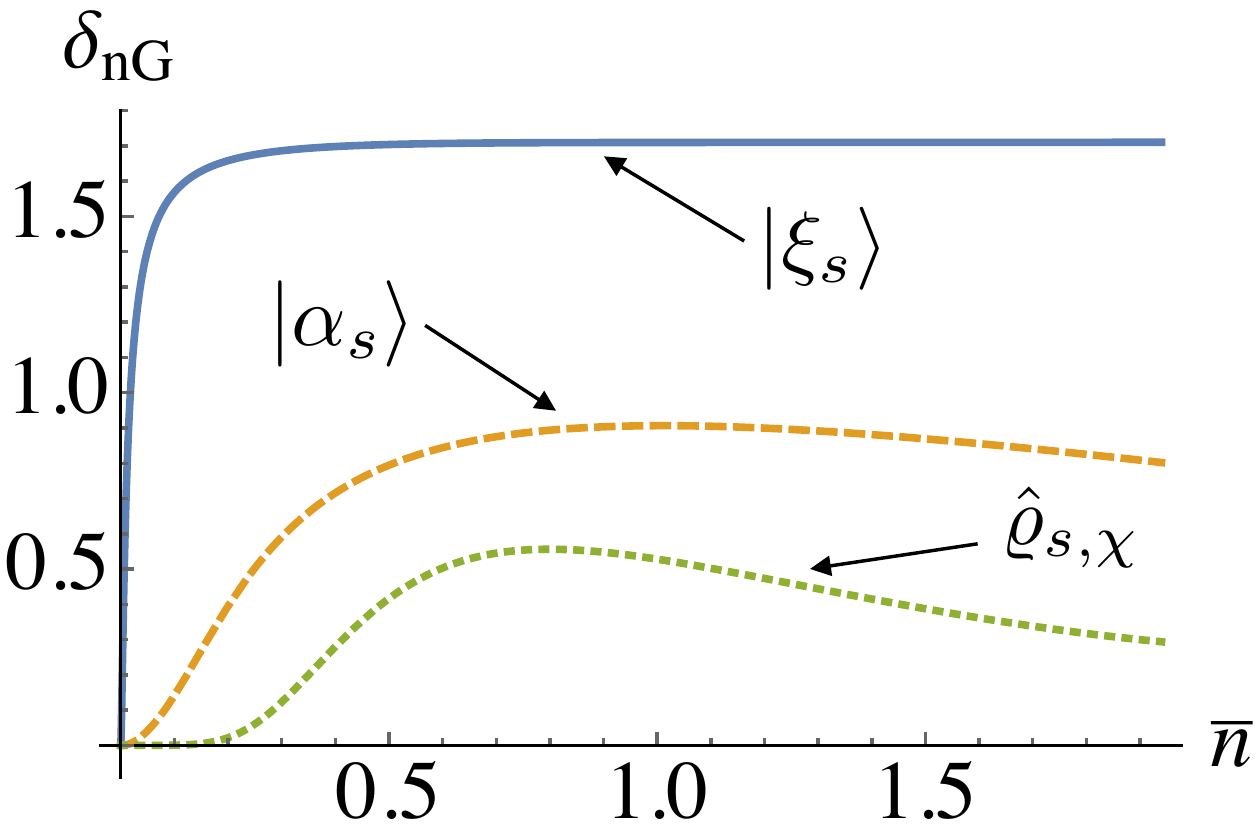}\includegraphics[width=0.245\textwidth]{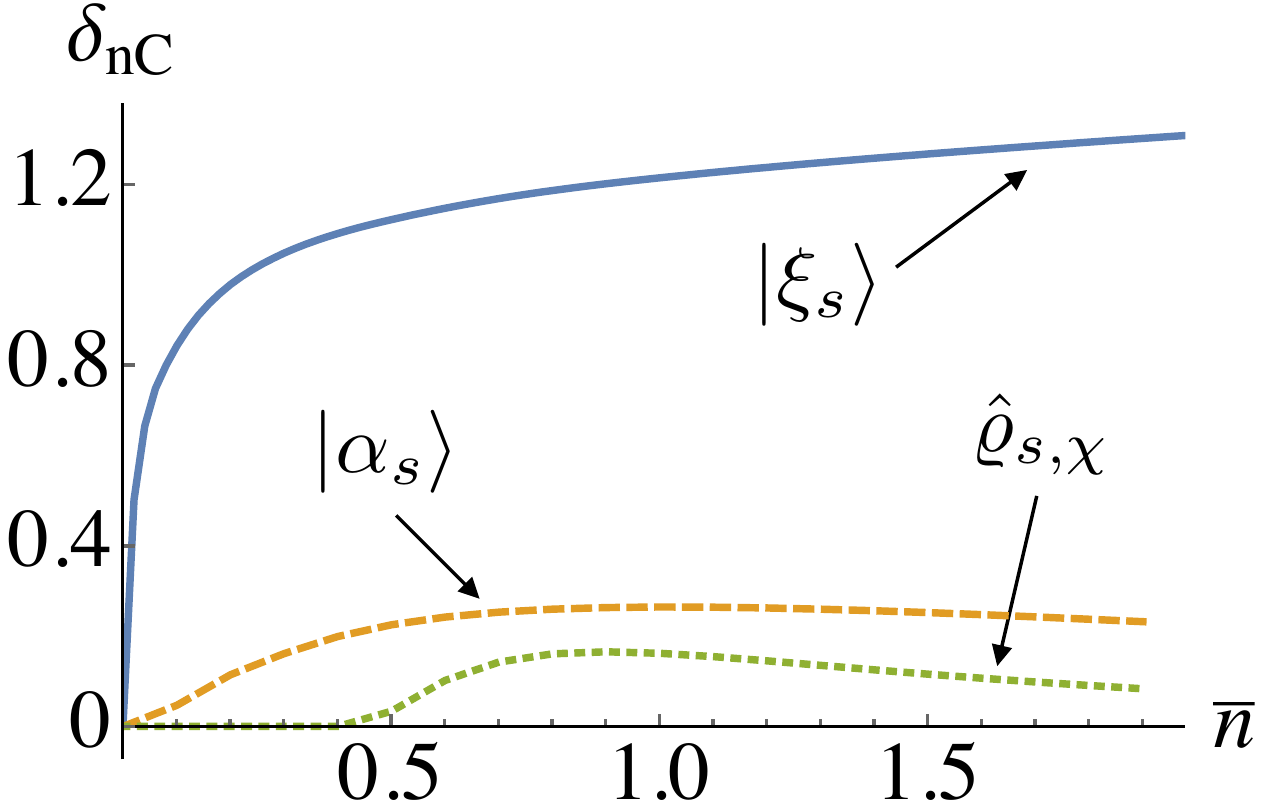}
	\caption{(Color online) Plots of the nG and nC measures, 
	$ \delta_\mathrm{nG} $ (left panel) and $\delta_\mathrm{nC}$ (right panel), 
	as a function of the mean input energy $\bar{n}$, for the 
	amplified states $\ket{\alpha_s}$ (dashed orange), $\ket{\xi_s}$ 
	(solid blue) and $\hat{\varrho}_{s,\chi}$ (dotted green). 
	The parameters of the NLA, $ g=3$ and $p=2 $, are fixed in 
	each plot.} \label{f:nGandnCCohvsSq}
\end{figure}

Let us now compare the performances of the amplification process 
in the generation of single-mode nG and nC.  To this purpose, we 
consider input signals with the same initial energy 
undergoing identical amplification process, i.e. 
$\bar{n}_\alpha=\bar{n}_\xi=\bar{n}_\chi\equiv \bar{n}$ where
$\bar{n}_\alpha\equiv \langle\alpha | \hat{a}^\dag \hat{a}|\alpha 
\rangle=|\alpha|^2$, $\bar{n}_\xi\equiv \langle\xi | \hat{a}^\dag 
\hat{a}|\xi \rangle=\sinh^2r$ and $\bar{n}_\chi=\chi(1-\chi^2)^{-1/2}$.
In Fig.~\ref{f:nGandnCCohvsSq} we plot $\delta_\mathrm{nG}$ (left panel) 
and $\delta_\mathrm{nC}$ (right panel) as a function of $\bar{n}$, with 
fixed NLA parameters $g=3$ and $p=2$. In particular, we see that 
for both the nG and nC measures the amplified squeezed vacuum
state (\ref{SqVacAmp}) performs better than the amplified coherent 
state (\ref{Coherent_s}), whereas the amplified reduced TWB state 
(\ref{TWBredAmp}) has the lowest values of nG and W-nC. We observe 
that the single-mode squeezing is a better resource in obtaining 
highly non-Gaussian and non-classical states, whereas the destructive 
measurement performed on the amplified TWB state somehow 
spoils the input quantumness.
%%% 
%%%%
\subsection{Increasing entanglement by non-destructive NLA on twin-beam}
\begin{figure}[h!]
\includegraphics[width=0.235\textwidth]{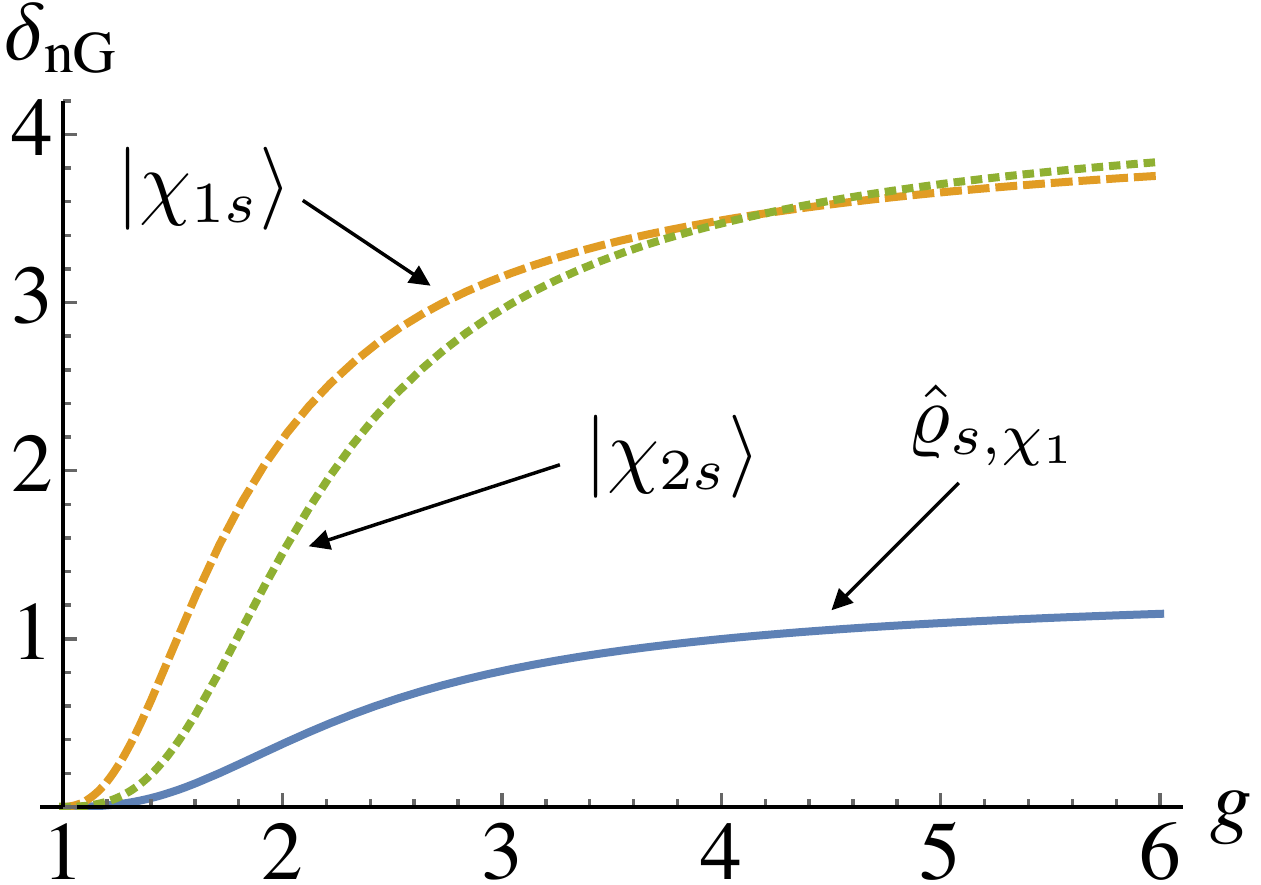}
\includegraphics[width=0.235\textwidth]{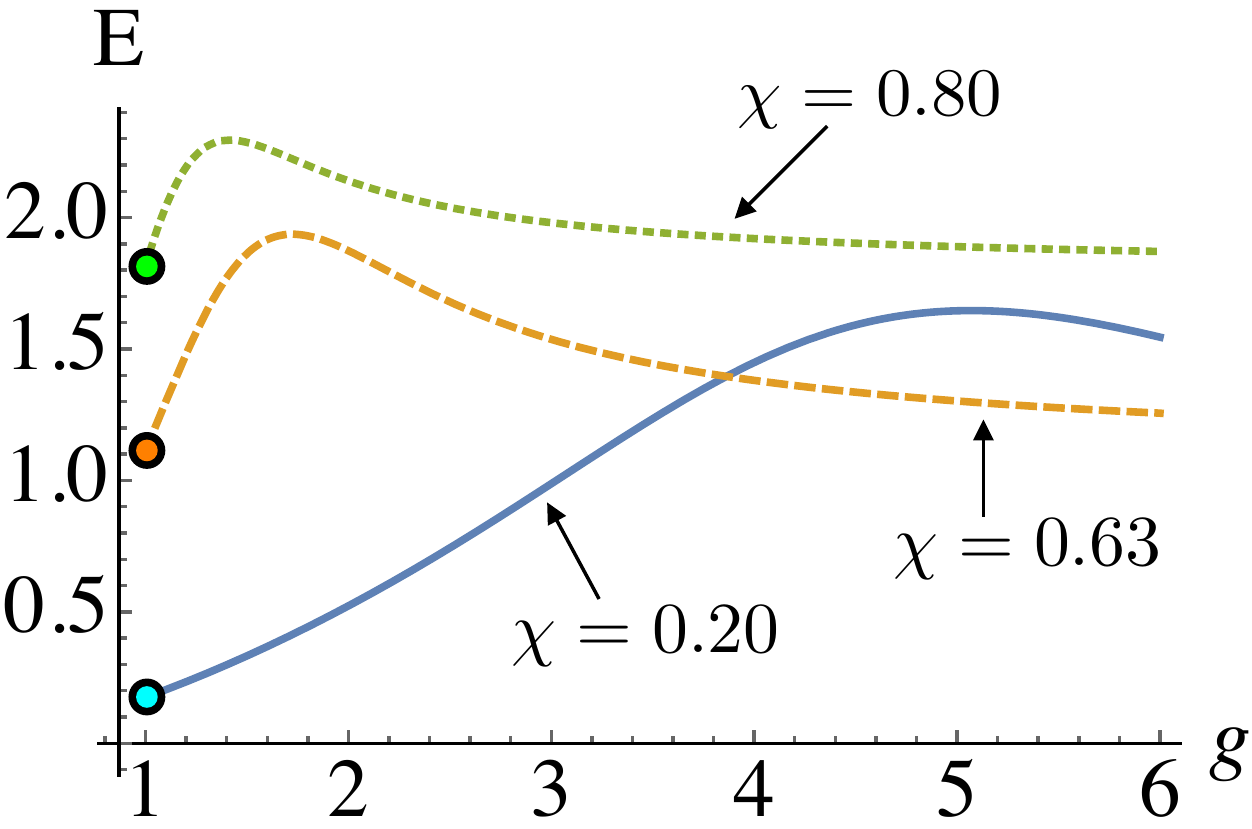}
\caption{(Color online) Left panel: nG measure $\delta_\mathrm{nG}$ plotted as a function of the gain parameter $g$ for the amplified states $\hat{\varrho}_{s,\chi_1}$ (solid blue curve), $\ket{\chi_{1s}}$ and $\ket{\chi_{2s}}$ (dashed orange and dotted green curves), with $\chi_1=0.63$ and $\chi_2= 0.50$. Right panel: Plot of entropy of entanglement $E[\ketbra{\chi}{\chi}]$ as a function of the gain parameter $g$. The three curves correspond to different squeezing parameters of the TWB state: $\chi=0.20$ (solid blue), $\chi=0.63$ (dashed orange), $\chi=0.80$ (dotted green). The three points at $g=1$ are the reference values of the entropy of entanglement for three initial TWB states corresponding to the three values of $\chi$ listed above.} \label{f:nGEntanglement}
\end{figure}

In this section, we examine the action of a successful non-destructive 
probabilistic amplification on a TWB state by focussing on the nG and 
the entanglement between the two correlated modes. A non-destructive 
measurement performed by the NLA on one mode of a TWB results in the
 following pure conditional state:
\begin{align}
\label{TWBamp}
\left\vert \chi_{s}\right\rangle = & \frac{\hat{E}_{s}^{p}\otimes\mathbb{I}_b\left\vert
	\chi\right\rangle}{\sqrt{P_{s,\chi}}} \\
=&\frac{\sqrt{1-\chi^{2}}}{\sqrt{P_{s,\chi}}}\left(  g^{-p}{\displaystyle\sum\limits_{n=0}^{p}}(g\chi)^{n}\left\vert n , n\right\rangle+
{\displaystyle\sum\limits_{n=p+1}^{\infty}}\chi^{n}\left\vert 
n,n\right\rangle \right) \, ,\notag 
\end{align}
where the success probability is given by Eq.~(\ref{Pschi}).
The nG measure for the amplified TWB state reduces to Eq.~(\ref{nGpure}), as the state (\ref{TWBamp}) remains pure under the action of the NLA. In order to compute the reference Gaussian state, we make some considerations on the CM (\ref{CM}) of a two-mode Gaussian state $\boldsymbol{\sigma}_2$, which can be written in a block form as
\begin{equation}
\boldsymbol{\sigma}_2=\left (\begin{array}{cc}
\boldsymbol{A} & \boldsymbol{C} \\
\boldsymbol{C}^T & \boldsymbol{B}
\end{array} \right ) \, ,
\end{equation}
where $\boldsymbol{A}$, $\boldsymbol{B}$ and $\boldsymbol{C}$ are $2\times 2$ matrices. By means of local symplectic transformations, it is possible to derive four symplectic invariants, namely $I_1\equiv \text{det}\,\boldsymbol{A}$, $I_2\equiv \text{det}\,\boldsymbol{B}$, $I_3\equiv \text{det}\,\boldsymbol{C}$ and $I_4\equiv \text{det}\,\boldsymbol{\sigma}_2$ and derive a simple expression for the two symplectic eigenvalues of  $\boldsymbol{\sigma}_2$:
\begin{equation}
d_\pm=\sqrt{\frac{\Delta(\boldsymbol{\sigma}_2)\pm\sqrt{\Delta^2(\boldsymbol{\sigma}_2)-4I_4}}{2}} \, ,
\end{equation}
where $\Delta(\boldsymbol{\sigma}_2)\equiv I_1+I_2+2I_3$. The von Neumann entropy of a generic two-mode Gaussian state can be written as $S(\hat{\varrho})=h(d_+)+h(d_-)$. In the case of the amplified TWB state (\ref{TWBamp}) it is easy to see that $I_1=I_2=\frac12 +\bar{N}_\chi$, with $\bar{N}_\chi=\text{Tr}[\ketbra{\chi}{\chi}(\hat{a}^\dag\hat{a}+\hat{b}^\dag\hat{b})]$, and that $d_+=d_-=\sqrt{I_1+I_3}$. 
Overall, the nG measure may be expressed as
\begin{equation}
\delta_\mathrm{nG}\Big[\ketbra{\chi_s}{\chi_s}\Big]=2 h(d_+)=2 \sqrt{I_1+I_3} \, 
\end{equation}
and it is plotted as a function of the gain parameter in the left panel of Fig.~\ref{f:nGEntanglement}. 
\par
As it is apparent from the plot, we recover the same behavior as in the previous examples concerning the monotonic growth with increasing gain of NLA. We compare the nG measure of the amplified states (\ref{TWBredAmp}) and (\ref{TWBamp}) for a NLA with, respectively, destructive and non-destructive measurements. In particular, by fixing the same squeezing parameter $\chi_1=0.63$ at $g=1$, meaning that the resource TWB state (\ref{TWB}) is fixed, we notice an enhancement of the NLA protocol in the non-destructive-measurement case (dashed curve with respect to the solid one). On the other hand, another comparison can be made by limiting the amount of mean energy of the resource state at $g=1$, i.e. fixing $\bar{n}_{\chi_1}=\bar{N}_{\chi_2}$ with $\chi_2= 0.50$. Also in this case, the enhancement of the amplification process occurs for a NLA with non-destructive measurement (dotted curve with respect to the solid one). The initial resources being equal, a NLA with non-destructive measurement acting on a TWB state strongly enhances the non-Gaussianity character of the amplified state.
\par
Let us now study whether the NLA enhances entanglement in the amplified 
TWB state. The most notable measure of entanglement for a bipartite pure state $\ket{\Psi}$ is the entropy of entanglement, which is defined as the von Neumann entropy of one of the reduced states $\varrho_b=\text{Tr}_a[\ketbra{\Psi}{\Psi}]$:
\begin{equation}
E[ \,\ket{\Psi}\,]  =S\left[  \hat{\varrho}_{b}\right] = -\sum_{n=0}^{\infty}\rho_{n}\ln \rho_{n} \, , \label{EntMeas}
\end{equation}
where $\rho_n$ are the eigenvalues of the reduced state. In our case, the initial TWB state is pure and the corresponding reduced state of mode $b$ is exactly the amplified state (\ref{TWBredAmp}) with a non-destructive measurement. We already calculated $S(\hat{\varrho}_{s,\chi})$ in the evaluation of the nG measure $\delta_\mathrm{nG}[\hat{\varrho}_{s,\chi}]$, highlighting the strong relation between non-Gaussianity of the reduced amplified state $\hat{\varrho}_{s,\chi}$ and the amount of entanglement of the bipartite amplified state $\ket{\chi_s}$. We can observe in the right panel of Fig.~\ref{f:nGEntanglement} that the entanglement measure $E[\,\ket{\chi_s}\,]$ shows a characteristic peak depending on the value of the squeezing parameter, occurring at lower values of the NLA gain for higher values of $\chi$. The remarkable result is provided by the comparison between the curves and the reference values of the amount of entanglement for the initial TWB state at $g=1$ (highlighted points in the figure). For any value of the NLA gain $g>1$, the amplification process brings along an enhancement of entanglement in the amplified state with respect to the amount of entanglement of the corresponding TWB state.
\section{Conclusion}\label{s:out}
In this work we have investigated the action of a non-deterministic 
noiseless linear amplifier on single- and two-mode Gaussian states 
in order to generate highly non-Gaussian and non-classical amplified 
quantum states. In particular, we have focused on amplification of feasible 
Gaussian states, e.g. coherent states, squeezed vacuum states and 
entangled twin-beam. 
\par
Our results show that noiseless amplification is, in general, a 
powerful scheme to generate non-classical non-Gaussian states, 
with the detailed performances depending on the interplay between 
the gain of the NLA, its threshold for amplification, and the 
parameters of the input signal. Upon comparing results for input 
signals with the same initial energy, we have shown that better 
performances, i.e. larger output nG and nC nG, are obtained by
amplification of squeezed vacuum. We have also analysed the 
performances of non-destructive amplification, showing that 
amplification of twin-beam highly increases entanglement.
\par
Concerning efficiency of the process, we have shown that
while the probability of successful amplification decreases 
with the NLA gain, there is a convenient trade-off between 
nG (or nC) and the success probability itself. We have also
shown that for squeezed vacuum input, one may increase nG and nC
at fixed success probability by increasing the NLA amplification 
threshold.
\par
Overall, we have proved that NLA provides an effective scheme 
to generate highly non-Gaussian and non-classical states and may 
be used to increase entanglement in continuous-variable systems.
Our results pave the way for optimised implementations of NLA and
suggest that both destructive and non-destructive 
implementations of NLA would be of interest in quantum technology.
\section*{Acknowledgements}
This work has been supported by SERB through project  VJR/2017/000011 
and by JSPS through project FY2017-S17118. MGAP is member of GNFM-INdAM.
FA acknowledges support from the UK National Quantum Technologies 
Programm (EP/M013243/1). The authors thank Luigi Seveso for several 
useful discussions.
\bibliography{NLAbib}
\end{document}